\documentclass[aps,prb,twocolumn,superscriptaddress,showpacs,floatfix,longbibliography]{revtex4-1}
\usepackage{amsmath,amssymb}

\usepackage[utf8]{inputenc}
\usepackage[T1]{fontenc}
\usepackage{xcolor}
\usepackage{booktabs}
\usepackage{array}
\IfFileExists{newtxtext.sty}
{\usepackage{newtxtext,newtxmath}}
{\IfFileExists{stix.sty}
	{\usepackage{stix}}
	{\IfFileExists{mathptmx.sty}
		{\usepackage{mathptmx}}{} } }

\usepackage{textcomp}

\usepackage{bm}

\IfFileExists{siunitx.sty}{\usepackage{booktabs,siunitx}}{}

\pdfoutput=1
\usepackage{color}
\definecolor{LinkColor}{rgb}{0.256,0.439,0.588}
\usepackage{hyperref}
\hypersetup{
	pdfauthor={abc},
	pdftitle={paper},
	colorlinks=true,
	citecolor=blue,
	linkcolor=blue,
	urlcolor=blue
}

\usepackage{pifont}

\usepackage[pdftex]{graphicx}
\DeclareGraphicsExtensions{.pdf,.jpeg,.png,.eps}
\usepackage{epstopdf}
\usepackage{soul}
\usepackage{lineno}
\begin{document}
	\bibliographystyle{unsrt}
	\title{Demystifying strange metal and violation of Luttinger theorem in a doped Mott insulator}
	
	\author{Wei-Wei Yang}
	\affiliation{School of Physical Science and Technology $\&$ Key Laboratory for Magnetism and Magnetic Materials of the MoE, Lanzhou University, Lanzhou 730000, People Republic of China} %

	\author{Yin Zhong}
	\email{zhongy@lzu.edu.cn}
	\affiliation{School of Physical Science and Technology $\&$ Key Laboratory for Magnetism and Magnetic Materials of the MoE, Lanzhou University, Lanzhou 730000, People Republic of China} %
	
	\author{Hong-Gang Luo}
	\affiliation{School of Physical Science and Technology $\&$ Key Laboratory for Magnetism and Magnetic Materials of the MoE, Lanzhou University, Lanzhou 730000, People Republic of China} %
	\affiliation{Beijing Computational Science Research Center, Beijing 100084, China}%
	
\begin{abstract}
Metallic states coined strange metal (SM), with robust linear-$T$ resistivity, have been widely observed in many quantum materials under strong electron correlation, ranging from high-$T_{c}$ cuprate superconductor, organic superconductor to twisted multilayer graphene and MoTe$_{2}$/WSe$_{2}$ superlattice. Despite decades of intensive studies, the mystery of strange metal still defies any sensible theoretical explanation and has been the key puzzle in modern condensed matter physics. Here, we solve a doped Mott insulator model, which unambiguously
exhibits SM phenomena accompanied with quantum critical scaling in observables, e.g. resistivity, susceptibility and specific heat. Closer look at SM reveals the breakdown of Landau's Fermi liquid without any symmetry-breaking, i.e. the violation of Luttinger theorem. Examining electron's self-energy extracted from numerical simulation provides the explanation on the origin of linear-$T$ resistivity and suggests that the long-overlooked static fluctuations in literature play an essential role in non-Fermi liquid behaviors in correlated electron systems.
\end{abstract}
	
	\date{\today}
	
	\maketitle
Metallic quantum states deviated from the prediction of Landau Fermi liquid (FL) theory have been widely observed in many quantum materials since the discovery of high-$T_{c}$ cuprate superconductor. Among these non-Fermi liquid (NFL) states, the strange metal (SM) with robust linear-$T$ resistivity stands out due to its ubiquitousness in correlated superconductors, heavy fermion compounds and recently uncovered twisted multilayer graphene.
Since in many cases of copper oxides the superconductivity emerges directly as an instability of the SM phase, it is believed that understanding the nature of SM should be the key step to solve the high-$T_{c}$ problem in cuprate and to establish the general framework for NFL phenomena \cite{RevModPhys.78.17,RevModPhys.83.1589,Powell_2011,Shen2020,daou2009linear,legros2019universal,kagawa2005unconventional,PhysRevLett.114.067002,PhysRevLett.50.270,bruin2013similarity,10.21468/SciPostPhys.6.5.061,keimer2015quantum,custers2003break}.

To attack SM, plenty of intriguing theoretical proposals are created wherein the aspect of doped Mott insulator (MI) due to Phil Anderson 
suggests a fruitful pathway culminated in classic $SU(2)$ gauge field theory. Unfortunately, the strong coupling nature of gauge theory hinders the correct solution of doped MI though the perturbative calculation does predict a linear-$T$ behavior in resistivity \cite{ANDERSON1196,PhysRevB.38.745,PhysRevLett.76.503}. Alternatively, recent progress on doped Hubbard model armed with state of art dynamic-mean-field-theory (DMFT) revisits the issue of doped MI, concluding that the elusive SM behaviors are caused by the doping-driven Mott quantum criticality \cite{PhysRevLett.114.246402,PhysRevB.88.075143}. More recent quantum Monte Carlo (QMC) study confirms SM at high-$T$ regime and suggests unambiguously the breakdown of Landau FL by the violation of Luttinger theorem \cite{landau1957theory,osborne2020fermisurface,huang2019strange}.

\begin{figure}[htp!]
	\centering
	\includegraphics[width=0.95\columnwidth]{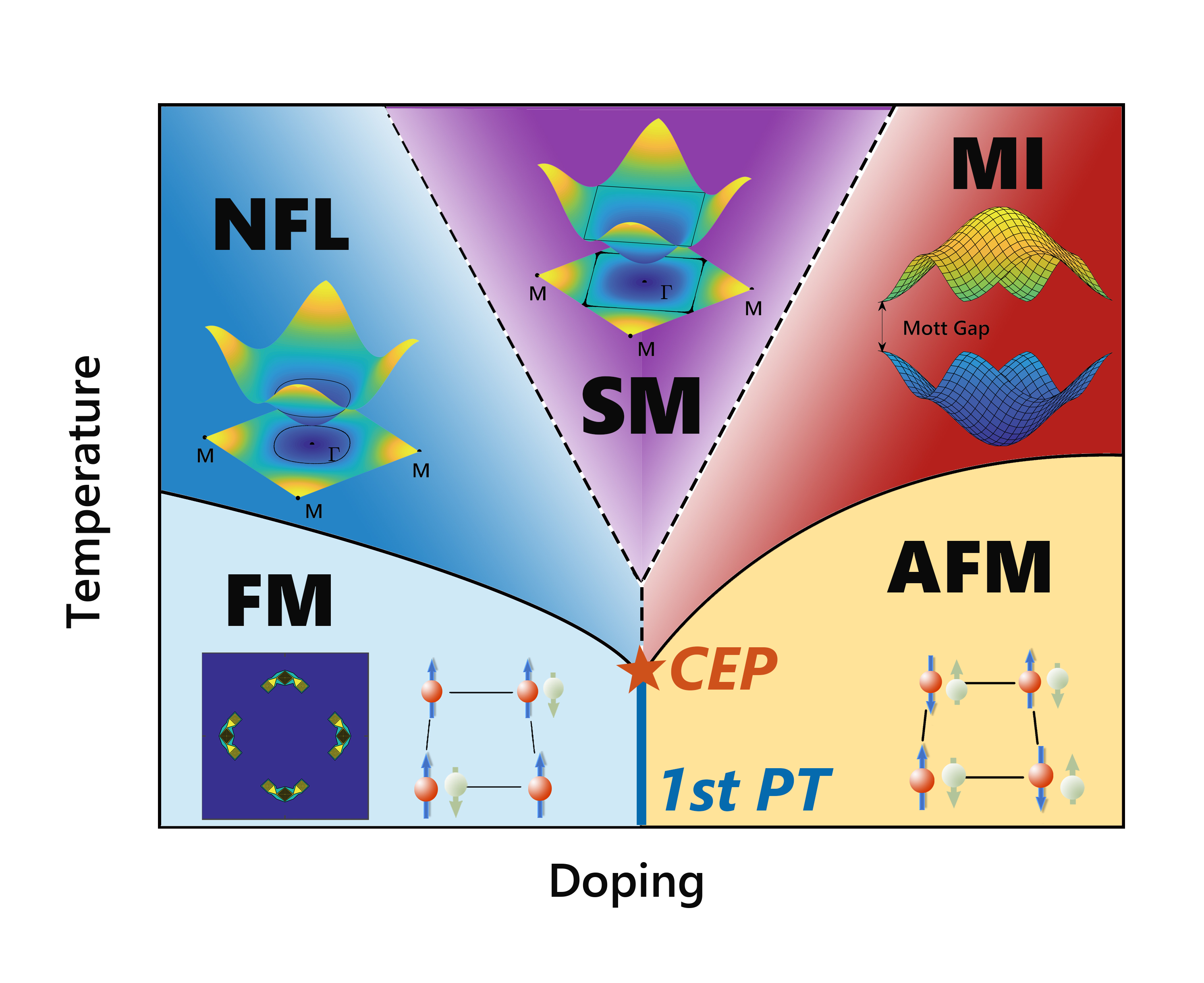}
	\caption{\textbf{Schematic phase diagram of the hole-doped Ising-Kondo lattice (IKL) model.} The low temperature regime is divided into an antiferromagnetic insulator (AFMI) and a ferromagnetic metal (FMM) by a first-order transition, which transforms to a weak first-order transition and finally ends at the critical end point (CEP). At high temperature above CEP, a quantum critical region (QCR) is induced by doping-driven Mott insulator-metal transition, which displays the strange metal (SM) behaviors. Out of QCR, there exists a Mott insulator (MI) at lower doping and a non-Fermi liquid (NFL) at a higher doping. Inset in MI shows two-band structure active at strong coupling while only lower band is depicted for SM and NFL.}
	\label{fig:1}
\end{figure}

\begin{figure*}[tp!]
	\centering
	\includegraphics[width=1.9\columnwidth]{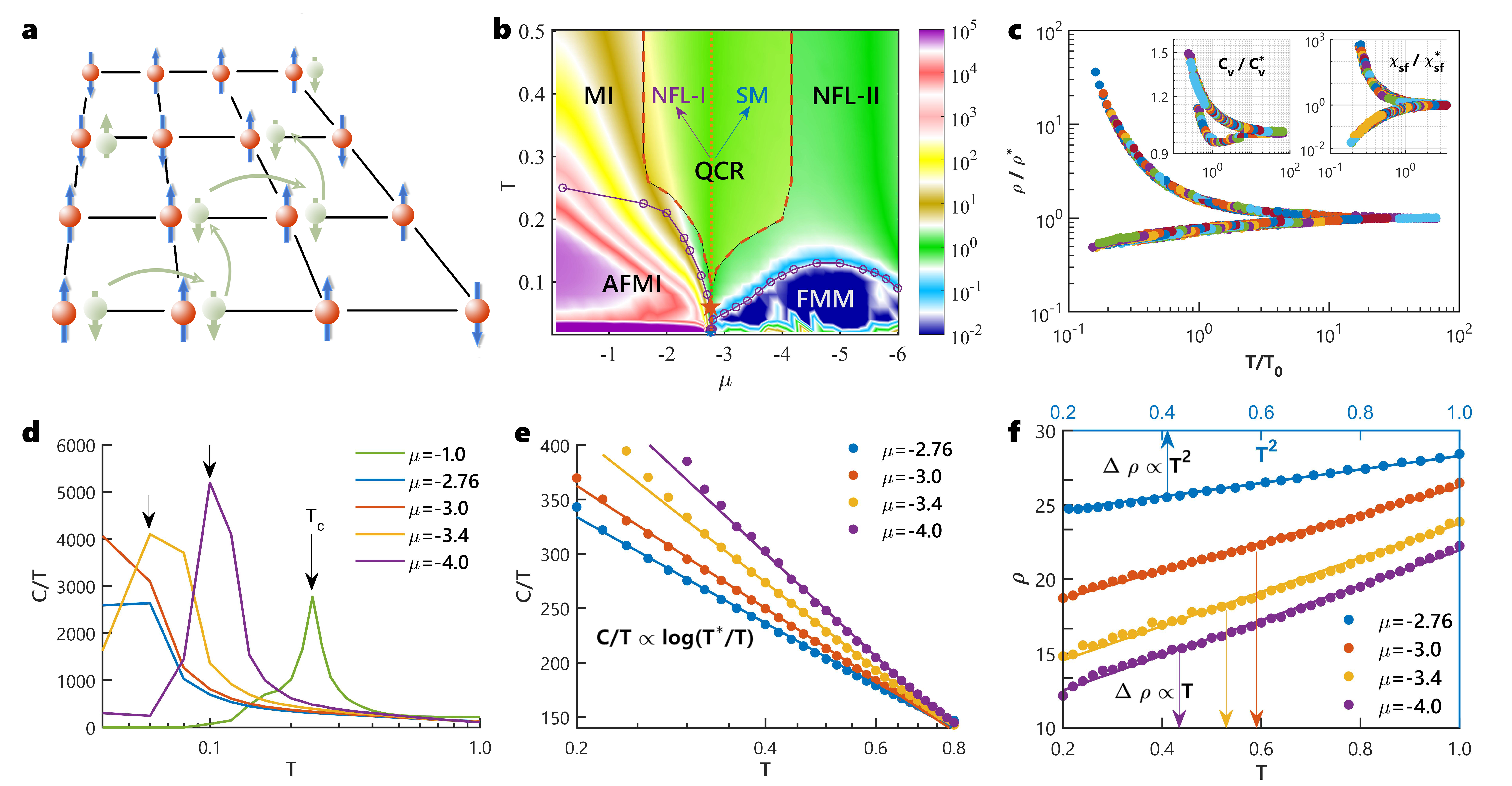}
	\caption{\textbf{Schematic picture, phase diagram of IKL and strange-metal behaviors in the QCR.} \textbf{a} Schematic representation of the doped IKL. \textbf{b} Phase diagram of the doped IKL model on square lattice in $\mu-T$ plane with $J=14$. Rich quantum states include AFMI, FMM, MI, insulator-like non-Fermi liquid (NFL-I), SM, and the second non-Fermi liquid (NFL-II). The background is a color map of the resistivity. Thermal fluctuation driven magnetic-paramagnetic phase transition is denoted by the purple line with circle mark. The first-order phase transition at ground state is marked by a blue circle. The CEP is marked by a red pentagram. The QCR is delimited by the crossover temperature (red dashed line). \textbf{c} Unconventional quantum scaling behavior of resistivity in the QCR. Family of resistivity curves are calculated along lines parallel to the separatrix $\mu^{\ast}(T)$ and all can be rescaled in the formula $\rho(T,\delta \mu)/\rho^{\ast}(T)=f(T/T_0(\delta \mu))$.  Inset plots the quantum scaling behavior of thermodynamic properties, i.e., heat capacity $C_v$ of $c$ electron and the susceptibility of $f$ electron $\chi_{sf}$.  \textbf{d-f}  Anomalous transport and thermodynamics in the SM. \textbf{d} Specific heat under different doping. The peak position corresponds to $T_c$ of magnetic-paramagnetic phase transition. \textbf{e} The same plot of specific heat in paramagnetic region with the $y$-axis magnified. In SM, $C(T)/T$ exhibits logarithm temperature dependence. The solid line is fitted by a $\texttt{log}(T^{\ast}/T)$ function. \textbf{f} Resistivity under different doping. In the QCR $\rho(T)$ satisfies a power-law dependence of temperature ($\rho(T)=\rho_0+AT^n$). With increasing doping, the resistivity changes from $T^2$-dependent behavior ($\mu=-2.76$) to a linear-$T$ dependence ($\mu=-3.0, -3.4, -4.0$), indicating the SM behavior.}
	\label{fig:2}
\end{figure*}

However, we have to emphasize that the mystery of SM has not been demystified due to the following reasons:(1) Since non-local spatial fluctuation which is important in quantum criticality has been largely neglected in DMFT, the Mott criticality driven SM phenomena may be fragile when spatial fluctuation are recovered; (2) The QMC study only establishes the existence of SM at rather high-$T$ while the fate of SM at physically more relevant low-$T$ regime (relevant to realistic experiments) has not been clarified due to the notorious fermions minus-sign problem for QMC;
(3) Only Hubbard model has been carefully examined, but models beyond Hubbard are crucial to understand many-bands systems like heavy fermion compounds \cite{RevModPhys.68.13,RevModPhys.78.865,limelette2003universality}; (4) Finally, no analytic or semi-analytic results are available in those numerical simulations, which are just black boxes if one cannot extract analytic formalism.

Motivated by aforementioned important progresses and unsolved issues, in this work we seek a clear and complete understanding about SM behaviors in doped MI with the help of an alternative model, which allows for nearly complete analytical and numerical treatment.
Our departure is the doped Ising-Kondo lattice (IKL) model, which naturally mimics the strongly correlated $f$ electron materials CeIrSn \cite{PhysRevB.98.155147}, TmB$_{4}$ \cite{PhysRevB.95.205140}, hidden order compound URh$_{2}$Si$_{2}$ \cite{PhysRevB.54.9322} and at the same time could be simulated by the unbiased Monte Carlo methods. In this solid platform, the competition between local and itinerant tendency involves non-trivial SM physics. The macroscopic anomalies include the robust nonsaturating, $T$-linear resistivity and the logarithmic temperature dependence of specific-heat coefficient. 
Such exotic SM behaviors beyond the quasiparticle paradigm turn to be intrinsically connected with the quantum criticality, which has also been observed in the vicinity of the quantum critical point (QCP) in various quantum critical heavy electron materials.\cite{Shen2020}
The absence of quasiparticle picture is clearly indicated by a strong violation of Luttinger theorem throughout the phase diagram. Further study shows that the IKL model exhibits these SM characters as a direct consequence of a NFL like self-energy, which provides natural routes to linear resistivity in SM phase.



\textbf{Results}

\textit{Model and phase diagram.}-We consider the IKL model on square lattice whose Hamiltonian is defined as
	\begin{equation}
	\hat{H}=-t\sum_{i,j\sigma}\hat{c}_{i\sigma}^{\dag}\hat{c}_{j\sigma}+
	\frac{J}{2}\sum_{j\sigma}\hat{S}_{j}^z\sigma \hat{c}_{j\sigma}^{\dag}\hat{c}_{j\sigma}-
    \mu\sum_{j\sigma} \hat{c}_{j\sigma}^{\dag}\hat{c}_{j\sigma},
	\label{eq:model1}
	\end{equation}
where $\hat{c}_{j\sigma}^{\dag}(\hat{c}_{j\sigma})$ is $c$ electron's creation (annihilation) operator with spin $\sigma=\uparrow,\downarrow$ at site $j$. $\hat{S}_{j}^{z}$ denotes the localized moment of $f$ electron. $t$-term denotes the hopping integral and only nearest neighbor hoping is involved. The hole doping into the half-filled system ($\mu=0$) is realized by tuning the chemical potential $\mu$. $J$ is the longitudinal Kondo coupling.
Choosing eigenstates of $\hat{S}_{j}^{z}$ as basis, Eq.~(\ref{eq:model1}) reduces into an effective free fermion model \cite{PhysRevB.100.045148} under fixed background $\{q_{j}\}$, ($q_{j}=\pm1$, $\hat{S}_{j}^{z}|q_{j}\rangle=\frac{q_{j}}{2}|q_{j}\rangle$) thus permits a straightforward classical (lattice) Monte Carlo simulation \cite{SM_MC}.
	
 In IKL model, the competition between itinerant and local tendency leads to different magnetic order, and thus further conspire to construct rich quantum states, including antiferromagnetic insulator (AFMI), ferromagnetic metal (FMM) at low temperature and paramagnetic phases as MI, SM, NFL at high temperature \cite{SM2}. There exists a critical end point (CEP, see the red pentagram in Fig.~\ref{fig:2}b), below which a first-order phase transition separates the IKL system into AFMI and FMM. Above CEP, a QCR with unusual quantum scaling behavior is uncovered. Our main result about the doped IKL is summarized in the $\mu-T$ phase diagram (see Fig.~\ref{fig:2}b), where the color bar denotes resistivity scaled by the one in the separatrix line (dotted line) of QCR \cite{SM3}.

\textit{Strange metal and quantum critical scaling}-The presence of SM state is ascertained by anomalous thermal and transport properties \cite{daou2009linear,legros2019universal,Varma_2016,RevModPhys.92.031001}.
We plot the evolution of heat capacity $C(T)$ and resistivity $\rho(T)$ with varying doping in Fig.~\ref{fig:2}d,e,f.
In SM phase, the heat capacity coefficient $C(T)/T$ increases with decreasing temperature, exhibiting an obvious dependence proportional to $\log(T^{\ast}/T)$ before approaching the magnetic phase transition, where $T^{\ast}$ is the cut-off temperature. As to resistivity, an obvious linear-$T$ dependence is revealed, which is nonsaturating with increasing temperature.

It has been suggested that it is the quantum critical physics who leads to both linear-$T$ resistivity and high-$T_c$ superconductivity. In most heavy fermion materials, the linear-$T$ resistivity is observed when tuning some external parameters to create the QCP \cite{hayes2016scaling,PhysRevLett.63.1996,kivelson1998electronic,van2003quantum,zaanen2004temperature}. 
In our model, even without the QCP, canonical signatures of quantum criticality are uncovered in a fan-like region together with the SM phase in various physical properties, including the resistivity, heat capacity and the magnetic susceptibility.
In Fig.~\ref{fig:2}c we plot the unconventional quantum scaling behavior of these transport and thermodynamic observables \cite{SM4}. 
As a whole, resistivity in the QCR (see fig.~\ref{fig:2}f) demonstrates a power-law dependence of temperature, satisfying $\rho(T)=\rho_0+AT^n$, and the $n$ gradually changes from 2 to 1 as the crossover to SM is occurring. Detailed study indicates the resistivity of QCR satisfies the following quantum critical scaling
\begin{equation}
    \rho(T,\delta \mu)=\rho^{\ast}(T)f(T/T_0(\delta \mu)),
	\label{eq:model2}
\end{equation}
where $T_0(\delta \mu)=c|\delta \mu|^{z\nu}$, $\delta \mu=\mu-\mu^{\ast}(T)$ and $\mu^{\ast}(T)$ is the critical 'zero field' trajectory corresponding to the 'separatrix' line \cite{SM1}. $\rho^{\ast}(T)$ is calculated for $\mu=\mu^{\ast}(T)$ and $f(x)$ denotes the unknown scaling function.The resistivity on the separatrix is almost independent with temperature. Note that at high temperatures, the resistivity curves depend weakly on different parameter, while as temperature is reduced the critical line separates resistivity curves to two branches. Here, we refer the insulator-like branch  to the first NFL state (NFL-I) \cite{SM5} and the metallic one to SM, respectively. The resistivity curves display the bifurcating characteristic. Crossing the center line, there is a change in trend with varying temperature. Both branches display the power-law scaling $T_0(\delta \mu)=c|\delta \mu|^{z\nu}$ with the same exponents $z\nu=1.25$, and the $T_0(\delta \mu)$ vanishes as $\delta \mu\rightarrow 0$, indicating a quantum criticality instead of classical phase transition \cite{SM6}.

\begin{figure}
	\centering
	\includegraphics[width=0.95\columnwidth]{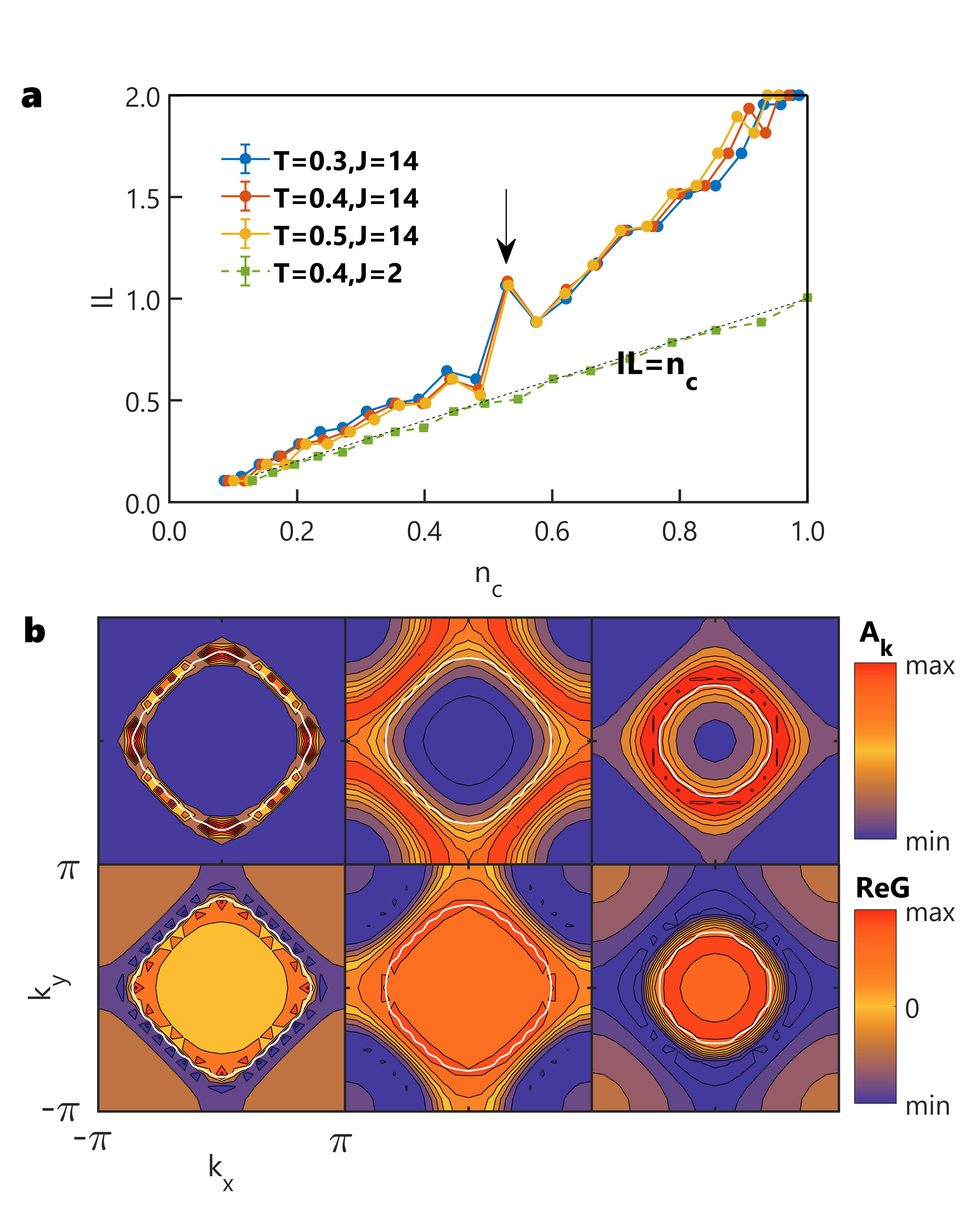}
	\caption{\textbf{Violation of Luttinger theorem.} \textbf{a} Luttinger integral (IL) versus density of electron $n_c$. In FL ($J=2$), the electron density is in accordance with the Luttinger theorem. At strong coupling ($J=14$), IL strongly deviates from the density of electrons for most doping regime thus confirms their NFL nature. IL diverges close to the boundary of QCR. Divergence is indicated by the arrow.  \textbf{b} Spectral function at Fermi energy $A_{\textit{\textbf{k}}}$$(\omega=0)$ (upper panel) and real part of Green function at Fermi energy $\mathrm{Re}G({\textit{\textbf{k}}},\omega=0)$ (lower panel) in different high-temperature phases: FL (left panel, $J=2$, $T=0.4$, $\mu=-0.8$); SM ($J=14$, $T=0.4$, $\mu=-3.0$, middle panel); NFL-II ($J=14$, $T=0.4$, $\mu=-4.6$, right panel). Electron density is denoted by the size of the Fermi surface's volume in a free system (white circle). The maximum of spectral function corresponds to the Fermi surface under strong coupling. }
	\label{fig:8}
\end{figure}

\textit{Violation of Luttinger theorem.}-Given the unconventional quantum criticality, it is natural to further confirm the quasi-particle properties of the exotic SM phase. To this aim, we examine the validity of Luttinger theorem in our model\cite{PhysRevB.68.085113,PhysRev.119.1153,PhysRevLett.84.3370}, 
which states that if Landau quasi-particle exists, the volume enclosed by Fermi surface is consistent with its density of particles.
Such theorem has been proved originally by Luttinger in terms of perturbation theory\cite{PhysRev.119.1153}, and later by Oshikawa's non-perturbation topological argument\cite{PhysRevLett.84.3370}. It now has been accepted as a key feature of FL. Mathematically, Luttinger theorem means the Luttinger integral ($\mathrm{IL}$) below must be equal to density of particles ($n_{c}$)
\begin{equation}
	\mathrm{IL}=\sum_{\sigma}\int_{\theta(\mathrm{Re}G(\textit{\textbf{k}},\omega=0))}\frac{d^dk}{(2\pi)^d}.
	\label{eq:model}
\end{equation}
However, we note the whole high temperature region of IKL violates the Luttinger theorem heavily.
The strong violation of the Luttinger theorem is demonstrated in Fig.~\ref{fig:8}a, which is robust with arbitrary doping for the strong coupling ($J=14$, $\mathrm{IL} \sim 2n_{c}$). It suggests a robust NFL-like nature for all paramagnetic phases in the phase diagram (see Fig.~\ref{fig:2}b), including SM. As a reference, we also show that the Luttinger theorem works well in the weak coupling case ($J=2$, $\mathrm{IL} \sim n_{c}$), agreeing with its FL nature. Extra feature in Fig.~\ref{fig:8}a is that the IL shows incipient divergence around $n_c \sim 0.53$, which is close to the boundary of QCR. 




The violation of Luttinger theorem could be characterized clearly by $c$ electron's single-particle spectral properties. As shown in Fig.~\ref{fig:8}b, $c$ electron's spectral function $A_{\textit{\textbf{k}}}(\omega=0)$ and the real part of Green function $\mathrm{Re}G(\textit{\textbf{k}},\omega=0)$ at Fermi energy are carefully studied. We compare the Luttinger integral and electron density in FL (left panel), SM (middle panel) and NFL-II (right panel). According to Luttinger theorem, electron density can be denoted by the size of the Fermi surface's volume in the free system (white line). Figure.\ref{fig:8}b shows nicely and clearly the working of the Luttinger theorem for the FL. Under weak coupling the volume enclosed by the Fermi surface is consistent with the one in free system ($n_c=0.36$). It suggests the presence of quasiparticle, since in this weak coupling situation the volume enclosed by the Fermi surface has not changed due to interaction. 
To the contrary, both the SM ($n_c=0.67$) and NFL-II ($n_c=0.31$) states display violation of Luttinger theorem, where the electron density is deviated from the volume enclosed by the Fermi surface, underlying the destruction of quasiparticle in the paramagnetic regime of the doped IKL.
\begin{figure*}[htp!]
	\centering
	\includegraphics[width=1.9\columnwidth]{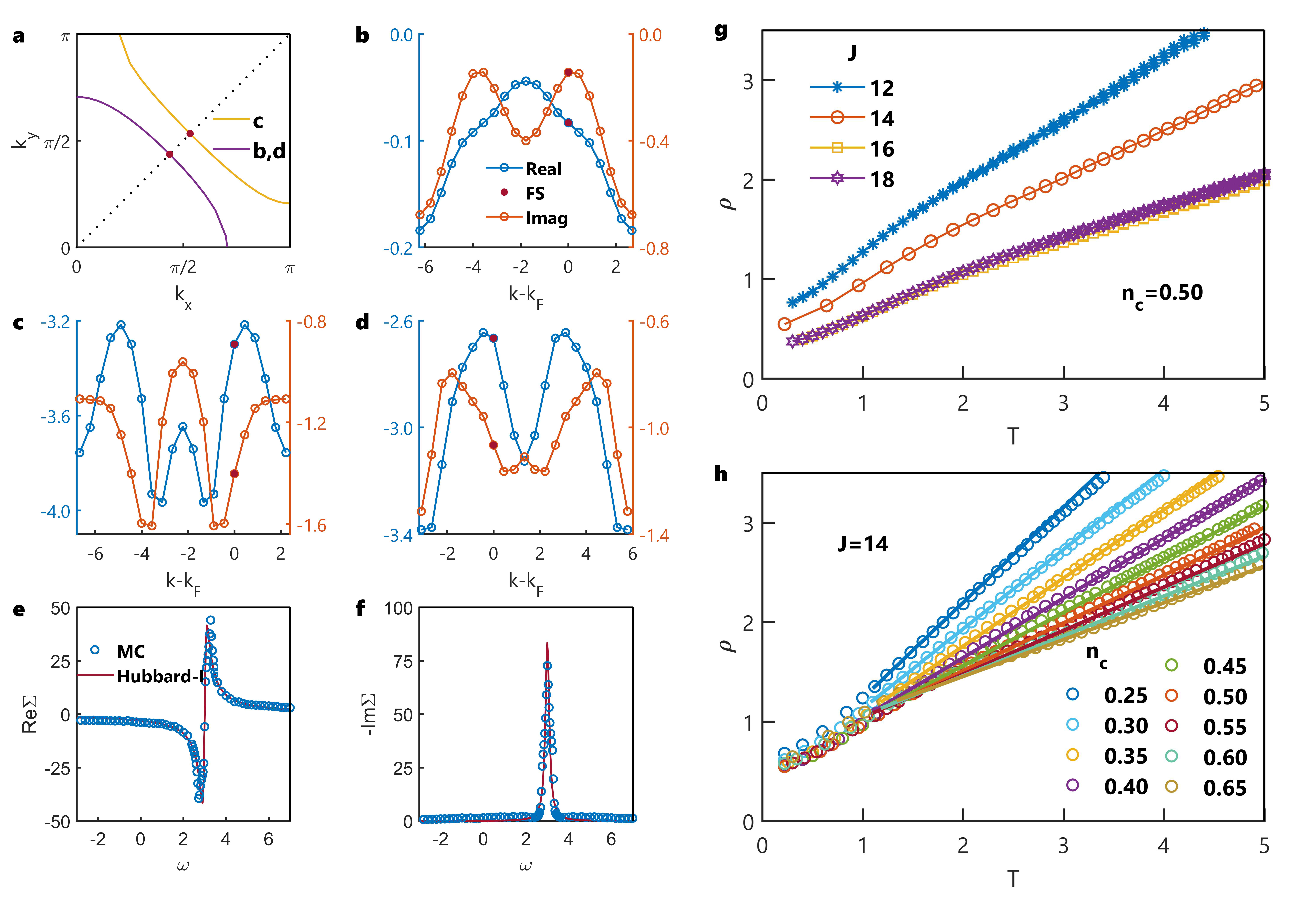}
	\caption{\textbf{NFL-like self-energy and the resultant linear-$T$ resistivity.} \textbf{a-d} Momentum dependence to the self-energy calculated around the Fermi energy. \textbf{a} Schematic picture for the Fermi surface in \textbf{b}-\textbf{d}. In \textbf{b} and \textbf{d}, Fermi surface (purple line) coincides with each other, whereas the electron density in \textbf{b} (FL, $\mathrm{IL}=0.72$) is double the one in \textbf{d} (NFL-II, $\mathrm{IL}=0.36$). The self-energy is calculated along the black doted line and the position of Fermi surface is marked with red circle. \textbf{b}-\textbf{d} Momentum dependence to the self-energy in FL (\textbf{b}, $J=2,T=0.4,\mu=-0.8$), SM (\textbf{c}, $J=14,T=1,\mu=-3.0$) and NFL-II (\textbf{d}, $J=14,T=1,\mu=-4.6$). In FL, the real part of the self-energy demonstrates a linear dependence of momentum, while the imaginary part shows a quadratic dependence, i.e., $\mathrm{Re} \Sigma (\omega=0) \sim k-k_\texttt{F}$ and $\mathrm{Im} \Sigma (\omega=0) \sim (k-k_\texttt{F})^2$. In contrast, the SM and NFL-II states display a linear momentum dependence for both the real and  imaginary parts of self-energy, $\mathrm{Re} \Sigma (\omega=0) \sim k-k_\texttt{F}$ and $\mathrm{Im} \Sigma (\omega=0) \sim k-k_\texttt{F}$. \textbf{e-f} Frequency dependence of self-energy in SM ($\mu=-3.0, T=1.0$). Result of the Hubbard-I approximation valid at strong coupling limit is displayed (red line), and it is in good agreement with MC data (blue circle). \textbf{g-h} Robust linear-$T$ resistivity revealed with eq.\ref{eq:se} at varying coupling strength within a wide range of doping. \textbf{g} Resistivity calculated under fixed doping ($n_c=0.50$) but varying coupling strength. \textbf{h} Resistivity calculated under different doping at fixed coupling strength $J=14$. The solid line is fitted by the function $\rho(T,n_c) \approx \rho_0 [n_c+\frac{C}{n_c}T]$.}
	\label{fig:w}
\end{figure*}

\textit{Microscopic mechanism underlying SM behaviors.}-Qualitatively, we have located the NFL nature of SM state by confirming the violation of Luttinger theorem. 
Next by extracting electron's self-energy $\Sigma (\textit{\textbf{k}},\omega)$ via Dyson equation, we can get more specific and concrete formula to describe these SM behaviors. 
We plot the real and imaginary parts of self-energy at Fermi energy respectively (see Fig.~\ref{fig:w}a-f). The Fermi surface is demonstrated in Fig.~\ref{fig:w}a and the momentum is chosen along the doted line. In FL, the quasiparticle dominates the low-energy physics and it leads to a quadratic-$k$ dependent behavior in imaginary part of self-energy around the Fermi energy due to the phase space effect. (As reference, we show the momentum dependence of FL ($J=2,T=0.4$) around Fermi surface in Fig.~\ref{fig:w}b. Its real part displays a linear dependence of $k$.)
Intriguingly, it turns out a quite different rule is enforced in the paramagnetic region of the IKL. In SM ($J=14, \mu=-3.0, T=1.0$) and NFL-II ($J=14, \mu=-4.6, T=1.0$), both real and imaginary parts demonstrate linear-$k$ dependence $\mathrm{Re}\Sigma(\omega=0)\sim k-k_\texttt{F}, \mathrm{Im }\Sigma(\omega=0)\sim k-k_\texttt{F}$.


Since the famous marginal FL theory has provided a reasonable functional form of self-energy to phenomelogically describe the normal state of the optimal doped cuprates \cite{PhysRevLett.63.1996},
a natural question posed by these observations is whether some quantitative microscopic description responsible for these SM behaviors is also accessible for the IKL system. 
Inspired by the analytic Hubbard-I approximation, we note the frequency dependence of self-energy can be fitted with $\Sigma (\omega) \sim \frac{J^2/16}{\omega+\mu+i\Gamma}$, as shown in Fig.~\ref{fig:w}e,f \cite{SM7}.
Thus in NFL-II/SM, the self-energy at the Fermi energy is approximately given by
\begin{equation}
\Sigma (\textit{\textbf{k}},\omega)=a\frac{J^2/16}{\omega+\mu+i\Gamma}+b(\textit{\textbf{k}}-\textit{\textbf{k}}_\texttt{F}),
\label{eq:se}
\end{equation}
where $\textit{\textit{\textbf{k}}}_\texttt{F}$ is the Fermi momentum.
Hitherto, several analytical approaches have been developed to study the source of linear-$T$ behavior, where the power-law functional form of self-energy ($\Sigma(\omega) \approx |\omega|^a$, where $a<1$) is the main goal \cite{PhysRevB.82.075127,PhysRevB.82.075128}. However, the exact form of self-energy can only be revealed in some soluble models, such as the Sachdev-Ye-Kitaev (SYK) model \cite{kitaev2015simple,PhysRevD.94.106002} and the Hatsugai–Kohmoto (HK) model \cite{Phillips_2020}. Equation.~\ref{eq:se} extracted from the reliable Monte Carlo simulation provides a distinct functional form of self-energy which could also result in the linear-$T$ resistivity. 

To further confirm the connection between SM state and the nontrivial self-energy formula, next we try to reproduce the linear-$T$ resistivity with Eq.~\ref{eq:se}.
We note this self-energy is quite robust under varying temperature at the whole paramagnetic region, which suggests a robust two-peak spectral function and at the same time a $T$-independent scattering rate. Previous study on the Hubbard and SYK models states that the broadening of spectral function by a $T$-independent scattering rate contributes heavily to the $T$-linear resistivity \cite{PhysRevResearch.2.033434}. Therefore, we expect the linear-$T$ resistivity emerges in the SM phase can as well be attributed to the $T$-independent feature of this non-trivial self-energy Eq.\ref{eq:se}.
With the optimal fitting coefficients \cite{SM9} of Eq.\ref{eq:se} we calculate the resistivity by the Kubo formula
\begin{equation}
	\sigma_{DC} =2 \pi \int d\epsilon \phi(\epsilon) \int \dfrac{\beta d \omega A(\epsilon,\omega)^2}{4\cosh^2(\beta\omega/2)}.
	\label{eq:Kubo}
\end{equation}
As shown in Fig.~\ref{fig:w}g,h, a strictly linear-$T$ dependence is demonstrated for all coupling strength, which is robust within a wide range of doping. Therefore, we conclude that our system accesses to robust linear-$T$ resistivity with the self-energy in the form of Eq.~\ref{eq:se}.

What's more, since the doping level determines the main energy scale, it also plays a dominant role in the transport properties. Unexpectedly, at most parameter region the slope of linear-$T$ resistivity is roughly proportional to the inverse of doping as shown in Fig.~\ref{fig:w}h, which could be fitted by
\begin{equation}
	\rho(T,n_c) \approx \rho_0 [n_c+\frac{C}{n_c}T] .
	\label{eq:rho}
\end{equation}
Importantly, such behavior is agreeing with recent experiments for cuprates, heavy fermion and quantum critical metal Sr$_{3}$Ru$_{2}$O$_{7}$ \cite{legros2019universal,bruin2013similarity} and also the magic-angle graphene \cite{PhysRevLett.124.076801}. This consistency underlines the basic physics captured by the IKL has shown the common trend of generic strongly correlated electron systems.

\textit{Discussion and conclusion.}-We remark that a similar phase diagram like Fig.~\ref{fig:2}b has been reported in the Hubbard model with the DMFT approximation \cite{PhysRevLett.114.246402,PhysRevB.88.075143}.
 Comparing the anomalous phenomena in the IKL and the Hubbard model, we find these two models share several common properties around Mott insulator-metal transition as following. (i) A robust NFL state is revealed around the transition with a linear-$T$ resistivity (e.g., Fig.~\ref{fig:2}f). (ii) The Mott transition is turned out to be third-order, indicated by the divergence of $\frac{\partial^2 n_c}{\partial \mu^2}$. (iii) Resistivity satisfies a quantum critical scaling like Eq.~\ref{eq:model2} with similar critical exponent ($(z\nu)_{\texttt{IKL}}=1.25$, $(z\nu)_{\texttt{Hubbard}}=1.35$).

There also exist some differences. The doped MI in the IKL has rigid band while in the Hubbard model spectral-weight transfer leads to non-rigid band.
Furthermore, our Monte Carlo simulation in the IKL includes both local and nonlocal correlation, and consequently the low temperature phases of the IKL turn out to be magnetic ordered states. Although at the absence of QCP, the quantum scaling behavior is confirmed in this solid platform. However, the DMFT applied in the Hubbard model makes a strong approximation such that the magnetic orders are suppressed at low temperature. 
Considering multiple low-temperature competing orders are all neglected, the uncovered QCP and even the quantum critical behavior in DMFT study might be artificial.

Comparing the IKL to solvable doped MI in the HK model,\cite{Phillips_2020} one see that no Luttinger surface exists in the IKL (also in the SYK and Hubbard) while the HK supports robust Luttinger surface characterized by zeros of Green function. Thus, the Luttinger surface may not be generic feature for doped MI though it does kill FL.

In conclusion, by studying a numerically solvable doped MI, exotic SM behaviors are revealed in the QCR, including the nonsaturating $T$-linear resistivity and the logarithmic temperature dependence of specific-heat coefficient.
These exotic behaviors elucidate the absence of quasiparticle, and it is further confirmed unambiguously by the violation of Luttinger theorem throughout the whole paramagnetic region in the phase diagram. All these SM characteristics all could be traced to an anomalous electron self-energy. The nontrivial formula for self-energy is extracted from the Monte Carlo data, which gives the robust linear-$T$ resistivity and the slope is consistent with the existing experimental data for several materials, e.g., the cuprates, heavy fermion and quantum critical metal Sr$_{3}$Ru$_{2}$O$_{7}$ \cite{legros2019universal,bruin2013similarity}, and even the magic-angle graphene \cite{PhysRevLett.124.076801}.
Compared with the classic Hubbard model, our study suggests that SM, Mott quantum criticality and the presence of Fermi surface with NFL-like self-energy are the intrinsic features of doped MI, which are promising to be revealed in generic strongly correlated electron systems.


\textbf{Methods}

Monte Carlo simulations of the IKL are carried out on the square lattice with periodic boundary conditions, where a $20 \times 20$ square lattice is mainly used. Accordingly, the two-dimensional Brillouin zone is sampled by a $20\times 20$ $k$-point grid. Nearest neighbor hopping integral is used as the unit ($t=1$) to measure all energy scales. To attack the Mottness, we focus on the strong coupling regime ($J=14$).

	\begin{acknowledgements}
This research was supported in part by Supercomputing Center of
Lanzhou University and NSFC under Grant No.~$11704166$, No.~$11834005$, No.~$11874188$, No.~$11674139$.	
We thank the Supercomputing Center of Lanzhou University for allocation of CPU time.		
	\end{acknowledgements}

\bibliography{main_ref}

\begin{thebibliography}{10}
\bibitem{SM6}
The critical exponent for $c$ electron ($(z\nu)_{c}=1.25$) is different from
  the one of $f$ electron ($(z\nu)_{f}=0.55$).
\end{thebibliography}




\end{document}